\documentclass{aa}  
\usepackage{natbib}
\bibpunct{(}{)}{;}{a}{}{,} 
\usepackage[normalem]{ulem}
\usepackage{xcolor}
\usepackage{siunitx}

\usepackage{graphicx}
\usepackage{txfonts}
\newcolumntype{P}[1]{>{\centering\arraybackslash}p{#1}}
\begin{document}

   \title{Evolution of the S\'ersic Index up to $z=2.5$ from JWST and HST}

   \author{M. Martorano\inst{1}
          \and
          A. van der Wel\inst{1}
          \and
          M. Baes \inst{1}
          \and
          E. F. Bell \inst{2}
          \and
          G. Brammer\inst{3}
          \and
          M. Franx \inst{4}
          \and
          A. Gebek \inst{1}
          \and
          S. E. Meidt \inst{1}
          \and
          T. B. Miller\inst{5}
          \and
          E. Nelson \inst{6}
          \and
          A. Nersesian \inst{7}$^,$\inst{1}
          \and
          S. H. Price \inst{8}
          \and
          P. van Dokkum \inst{9}
          \and
          K. E. Whitaker\inst{3}$^,$\inst{10}
          \and
          S. Wuyts \inst{11}
          }

   \institute{Sterrenkundig Observatorium, Universiteit Gent, Krijgslaan 281 S9, 9000 Gent, Belgium\\
              \email{marco.martorano@ugent.be}
         \and
             Department of Astronomy, University of Michigan, 1085 South University Avenue, Ann Arbor, MI 48109–1107, USA
         \and
             Cosmic Dawn Center (DAWN), Niels Bohr Institute, University of Copenhagen, Jagtvej 128, København N, DK-2200, Denmark
         \and
             Leiden Observatory, Leiden University, P.O. Box 9513, 2300 RA, Leiden, The Netherlands
         \and
            Center for Interdisciplinary Exploration and Research in Astrophysics (CIERA), Northwestern University, 1800 Sherman Ave, Evanston, IL 60201, USA
         \and
             Department for Astrophysical and Planetary Science, University of Colorado, Boulder, CO 80309, USA
         \and
             STAR Institute, Université de Liège, Quartier Agora, Allée du six Aout 19c, B-4000 Liege, Belgium
         \and
             Department of Physics and Astronomy and PITT PACC, University of Pittsburgh, Pittsburgh, PA 15260, USA
         \and
             Department of Astronomy, Yale University, 52 Hillhouse Avenue, New Haven, CT 06511, USA
         \and
             Department of Astronomy, University of Massachusetts, Amherst, MA 01003, USA
         \and
             Department of Physics, University of Bath, Claverton Down, Bath, BA2 7AY, UK
             }

   \date{Received November 07, 2024; accepted January 03, 2025}

 
  \abstract
   {The Hubble Space Telescope (HST) has long been the only instrument able to allow us to investigate the structure of galaxies up to redshift $z=3$, limited to the rest-frame UV and optical. The James Webb Space Telescope (JWST) is now unveiling the rest-frame near-IR structure of galaxies, less affected by dust attenuation and more representative of their underlying stellar mass profiles.}
   {We measure the evolution with redshift of the rest-frame optical and near-IR S\'ersic index ($n$), and examine the dependence on stellar mass and star-formation activity across the redshift range $0.5\leq z\leq2.5$.}
   {For an HST-selected parent sample in the CANDELS fields we infer rest-frame near-IR S\'ersic profiles for $\approx 15.000$ galaxies in publicly available NIRCam imaging mosaics from the COSMOS-Web and PRIMER surveys. We augment these with rest-frame optical S\'ersic indices, previously measured from HST imaging mosaics.} 
   {The median S\'ersic index evolves slowly or not at all with redshift, except for very high-mass galaxies ($M_\star > 10^{11}~{\text{M}}_\odot$), which show an increase from $n\approx 2.5$ to $n\approx 4$ at $z<1$.  High-mass galaxies have higher $n$ than lower-mass galaxies (the sample reaches down to $M_\star=10^{9.5}~{\text{M}}_\odot$) at all redshifts, with a stronger dependence in the rest-frame near-IR than in the rest-frame optical at $z>1$. This wavelength dependence is caused by star-forming galaxies that have lower optical than near-IR $n$ at z>1 (but not at z<1). Both at optical and near-IR wavelengths, star-forming galaxies have lower $n$ than quiescent galaxies, confirming and fortifying the result that across cosmic time a connection exists between star-formation activity and the radial stellar mass distribution. Besides these general trends that confirm previous results, two new trends emerge: 1) at $z>1$ the median near-IR $n$ varies strongly with star formation activity, but not with stellar mass, and 2) the scatter in near-IR $n$ is substantially higher in the green valley (0.25 dex) than on the star-forming sequence and among quiescent galaxies (0.18 dex) -- this trend is not seen in the optical because dust and young stars contribute to the variety in optical light profiles. Our newly measured rest-frame near-IR radial light profiles motivate future comparisons with radial stellar mass profiles of simulated galaxies as a stringent constraint on processes that govern galaxy formation.} 
   {}


   \maketitle
%

\section{Introduction}

The structure of a galaxy contains key information about the processes that determine its evolutionary history, such as accretion, merging, and secular processes. 
Morphology strongly relates to the star formation rate (SFR): young star-forming galaxies tend to have disks with exponential profiles while older, quiescent galaxies have higher average densities and steeper radial profiles, dominated by a bright center \citep[e.g.,][]{kauffmann03a,franx08, wuyts11, bell12}. These are usually associated with the presence of a bulge \citep{gadotti09, bluck14, salo15}, which, in turn, is related to the mass of the central black hole. The latter is invoked to be one of the mechanisms responsible for the cessation of star formation \citep{chen20, bluck23} via radio-mode feedback \citep{croton06}.

The physical cause driving the transition of a galaxy from the star-forming phase to the quenched phase is still under debate and it is yet unclear whether the morphological change in the galaxy's structure is causally connected to quenching directly or indirectly.
We know, however, that the presence of a steeper than exponential light profile is a necessary but not sufficient condition for a galaxy to be considered quiescent \citep[i.e.][]{bell12, lang14, whitaker17, martorano23}.
In other words, a non-negligible fraction of bulge-dominated galaxies is star-forming \citep[e.g.,][]{bell12, barro14}, suggesting that central densities build up before quenching.

Even though several methods exist to investigate galaxy structure (i.e., multi-component analysis, non-parametric methods, bulge-disk decomposition and many others; see \citealt{conselice14} for a complete review), the S\'ersic profile \citep[$\log(I(r))\propto r^{-n}$][]{sersic68} is the most used parameterization to describe radial light profiles with $n$ (the S\'ersic index) describing its radial shape (i.e. $n=1$ exponential-like and $n=4$ de Vaucouleurs-like profile). Its usage is aided by the availability of several public software packages such as \textsc{Galfit} \citep{peng02,peng10}, \textsc{Profit} \citep{robotham16} or \textsc{PySersic} \citep{pasha23} which implement it for light-profile fitting.

CANDELS \citep{grogin11, koekemoer11} is the largest survey in the rest-frame optical regime with sufficient spatial resolution and depth to allow for S\'ersic index measurements around cosmic noon. Although the dependence of S\'ersic index on SFR and stellar mass has been deeply investigated at low-z \citep[i.e.][]{lange15}, remarkably, just a few works \citep[i.e.][]{patel13, lang14, whitaker15} investigated the same dependencies across redshift taking advantage of this dataset even though S\'ersic profile measurements have been available for over a decade \citep{van-der-wel12}. \citet{lang14} study the mass dependence in two broad redshift bins ($0.5<z<1.5$ and $1.5<z<3$) finding that, at a fixed stellar mass, the S\'ersic index weakly changes with redshift but has a strong mass dependence for both star-forming and quiescent galaxies. On the other hand, with the same data, \citet{patel13} and \cite{van-dokkum13} demonstrate that individual galaxies must, as they increase in stellar mass, strongly increase their S\'ersic index over cosmic time.  Finally, \cite{shibuya15} showed that the S\'ersic index for star-forming galaxies, averaging over a broad range in stellar masses, remained nearly constant (at $n\sim1.4$) across the redshift range $z=0.5-2$. In short, there has been no detailed description of the joint stellar mass- and redshift-dependence of the S\'ersic index. This is one (of two) main motivations for this paper.

Until the launch of the James Webb Space Telescope (JWST), the only instrument with sufficient angular resolution to allow for the quantification of light profiles at high redshift was the Hubble Space Telescope (HST), limited to the rest-frame optical (at $z\lesssim 2.5$) or UV (at $z\gtrsim 2.5$). At these wavelengths, the outshining effect \citep{papovich01,maraston10, wuyts11, reddy12, sorba18, leja19b, suess22b, narayanan24} from young stellar populations and the dust attenuation affect the light distribution, exacerbating the difference between light and mass profiles. 
These issues are finally circumvented to a large extent by the wavelength coverage provided by JWST, producing rest-frame near-IR light profiles up to $z\sim3$. For example, recent studies combining radio, near-IR and mid-IR observations with ALMA, JWST/NIRCam and JWST/MIRI \citep[i.e.][]{chen22, tadaki23, lebail23} reveal significant obscured star formation in the center of galaxies. Several other pioneering works are taking advantage of the synergy between these instruments to address the near-IR structure of high-redshift galaxies \citep[e.g.,][]{suess22, price23, gillman23, gillman24, cutler24, ward24, costantin24, shivaei24}.
In \cite{martorano23} we show that for most galaxies at redshifts $0<z<3$, the S\'ersic index does not greatly differ between the rest-frame optical and rest-frame near-IR. The limited sample size of high-mass galaxies covered by the CEERS \citep{Finkelstein17, Finkelstein23} footprint, used for that work, limited the statistical power, making it hard to discern more subtle differences between rest-frame optical and near-IR S\'ersic indices. 
The second motivation of the current paper is to take advantage of the much larger area covered by JWST/NIRCam COSMOS-Web \citep{casey23} and PRIMER-COSMOS \citep{dunlop21} surveys and use samples of stellar mass-selected galaxies at $0.5\leq z\leq2.5$ with rest-frame near-IR imaging that rivals the CANDELS data set in sample size and spatial resolution.

The paper is structured as follows. In Section \ref{sec: Data} we describe the data sets used and the selection of the investigated galaxy sample. Section \ref{sec: results} is dedicated to the presentation and discussion of the correlations between S\'ersic index and stellar mass, redshift, and SFR. Finally, in Section \ref{sec: conclusion}, we summarize the content of the paper and draw our conclusions.

Throughout the paper we use the AB magnitude system \citep{oke83} and assume a standard Flat-$\Lambda$CDM model with $\Omega_{\text{m}}$=0.3 and $H_0 = 70~{\text{km}}~{\text{s}}^{-1}~{\text{Mpc}}^{-1}$.

\section{Data} \label{sec: Data}
    In this section, we present an overview of the data and the galaxy selection procedure to create a robust and consistent sample.
    
    \subsection{HST Dataset} \label{sec: candels data}
    We use cataloged HST data presented by \cite{van-der-wel12} based on HST observations obtained as part of the CANDELS program \citep{koekemoer11, grogin11}. These observations cover all five CANDELS fields (COSMOS, UDS, EGS, GOODS-South, and GOODS-North) in the two WFC3 filters F125W and F160W with a 5$\sigma$ depth of H$_{F160W}$=27ABmag. The catalog contains $186\,440$ sources whose morphological parameters have been recovered via S\'ersic profile fitting using the software package \textsc{Galfit} \citep{peng02,peng10}, as outlined in \cite{van-der-wel12}.  The uncertainties in the measured S\'ersic indices are not taken from the \textsc{Galfit} profile fits, but calculated from the signal-to-noise ratio, calibrated to account for the total random uncertainty, as outlined by \citealt{van-der-wel12}.
    
    Stellar mass, redshift and star formation rates are taken from the catalog presented by \cite{leja20}, who combine HST, Spitzer and ground-based photometric observations from the near-UV to 24$\mu$m \citep{skelton14, whitaker14} to perform spectral energy distribution (SED) fitting with the code \textsc{Prospector} \citep{Johnson17, Johnson21} for $63\,413$ galaxies. These are selected from 3D-HST \citep{brammer12} to have redshift between 0.5 and 3 and signal-to-noise ratio ${\text{SNR}}_{F160W}>10$. For $\sim5\%$ of the sample spectroscopic redshift is available, $\sim20\%$ has grism redshift and the remaining $\sim75\%$ of the sample has a photometric redshift. These redshift values are fixed during the SED fit.
    The \textsc{Prospector} models include nonparametric star formation histories (SFH), a Chabrier \citep{chabrier03} initial mass function (IMF), a two-component dust attenuation model with flexible attenuation curve, variable stellar metallicity and dust emission powered by energy balance. A detailed setup description is presented in \cite{leja19,leja20}.

    We cross-match the \cite{van-der-wel12} and \cite{leja20} catalogs, keeping sources with an angular separation below \SI{0.4}{\arcsecond} which leaves us with $60\,504$ galaxies.
    
    The rest-frame 0.5$\mu{\text{m}}$ S\'ersic indices ($n_{0.5\mu{\text{m}}}$) are calculated by linearly interpolating the cataloged values of $n$ in F125W and F160W, using the respective pivot wavelengths. This interpolation is repeated 100 times, sampling the measurements from the F125W and F160W $n$ uncertainties and taking the median values of the retrieved interpolation at 0.5$\mu{\text{m}}$ as the best $n_{0.5\mu{\text{m}}}$ estimate.
    
    We limit the redshift range to $0.5-2.5$ to avoid strong extrapolation effects, rejecting $3\,708$ galaxies.
    We set a stellar-mass threshold of $M_\star=10^{9.5}~{\text{M}}_\odot$ which grants mass completeness up to $z=2.5$ \citep{tal14}, remaining with $22\,963$ galaxies.
    
    The catalog provided by \cite{van-der-wel12} includes a \textit{"\textsc{Galfit} model quality flag"} that we use to reject $2\,265$ galaxies whose S\'ersic fit did not properly converge (flag$\geq$2). To increase the statistics, upon verifying this does not bias our sample, we decided to keep in the sample those $3\,578$ galaxies cataloged as suspicious fit (flag$=$1) in \cite{van-der-wel12}.

   Following the finding by \cite{van-der-wel12} that reliable S\'ersic index measurements requires ${\text{SNR}}_{F160W}\gtrsim50$, we set this condition on the ${\text{SNR}}$ in the filter closer to the rest-frame wavelength $0.5\mu{\text{m}}$. This reduces our sample to $14\,826$ galaxies.
    
    Among these, 208 have $n_{0.5\mu{\text{m}}}$ outside the 0.2-8 range which defines the fit constraints set by \cite{van-der-wel12}. For these galaxies, we set $n_{0.5\mu{\text{m}}}$ to the nearest boundary value ($n=0.2$~or $n=8$).   Removing these sources from the sample induces negligible variations in the results (variations on the medians are below $4\%$). These targets do not introduce a relevant bias.
    We separate quiescent and star-forming galaxies at $0.8$~dex below the star-forming main sequence (SFMS) ridge defined by \cite{leja22}, resulting in 
    $2\,638$ quiescent galaxies and 
    $12\,188$ star-forming galaxies.

    \subsection{JWST Dataset} \label{sec: jwst data}
    In \cite{martorano24} we investigate the size-mass distribution at rest-frame 1.5~$\mu$m for  $\sim26\,000$ galaxies in the COSMOS-Web \citep{casey23} and PRIMER-COSMOS \citep{dunlop21} fields observed with JWST/NIRCam in the redshift range $0.5-2.5$ and with stellar mass $M_\star>10^9~{\text{M}_\odot}$. In the present paper, we use the same \textsc{GalfitM} \citep{haussler13} S\'ersic profile fits of the F277W and F444W imaging presented in \cite{martorano24}.

    The parent sample used by \cite{martorano24} is drawn from the multi-wavelength COSMOS2020 catalog \citep{weaver22} after the exclusion of AGN candidates detected in \cite{chang17}. We pre-select our sample based on the cataloged \textsc{LePhare} \citep{arnouts02, ilbert06} stellar masses (M$_\star>10^{9.5}$~M$_\odot$) and redshift ($0.5\leq z\leq2.5$). The different stellar mass threshold adopted in this work makes us reject $10\,441$ galaxies.
    For consistency with the HST/CANDELS sample in the rest of the work we make use of stellar population parameters inferred with the code \textsc{Prospector}. Therefore, for all of the galaxies selected from \cite{martorano24}, we perform SED fits with \textsc{Prospector} using photometry from the COSMOS2020 catalog, fixing the redshift to the \textsc{LePhare} value, and adopting the same \textsc{Prospector} setup as \cite{leja20}.
    For 388 galaxies we retrieve a \textsc{Prospector} stellar mass below the threshold adopted, hence we remove them from the sample.

    We remind the reader that the HST/CANDELS dataset and the COSMOS2020 catalog (hence the catalog by \cite{martorano24} that is a subset of COSMOS2020) just partially overlap and only $\sim10\%$ of the JWST sample used in this work was observed also with HST during the CANDELS program.
    To check the consistency between our \textsc{Prospector} run and \cite{leja20} (based on COSMOS2015 \citep{laigle16} and 3D-HST \citep{brammer12} photometry), in Appendix \ref{appendix: dataset_comp} we compare the stellar masses and star-formation rates for the subsample of $1\,656$ galaxies in common between the two datasets finding a good agreement.

    The rest-frame 1.5$\mu{\text{m}}$ S\'ersic index ($n_{1.5\mu{\text{m}}}$) is retrieved from the catalog published in \cite{martorano24} where authors computed it in the same manner as for the HST sample (Sect.~\ref{sec: candels data}). Also for this sample we require the ${\text{SNR}}$ in the filter closest to the rest-frame wavelength of interest ($1.5\mu{\text{m}}$) to be larger than 50. This criterion rejects 248 galaxies.
    75 other galaxies have $n_{1.5\mu{\text{m}}}$ outside the $0.2-8$ range, hence, we set their $n_{1.5\mu{\text{m}}}$ to the nearest boundary value ($n=0.2$~or $n=8$).
    The final near-IR sample contains $14\,882$  galaxies, $11\,965$ of which are classified as star-forming and $2\,917$ as quiescent, adopting the selection based on the SFMS ridge defined by \cite{leja22}.

\section{Results and discussion}\label{sec: results}

\subsection{Mass and redshift dependence of the S\'ersic index}\label{sec:n-m-z}

    \begin{figure*}
        \centering
        \includegraphics[scale=0.5]{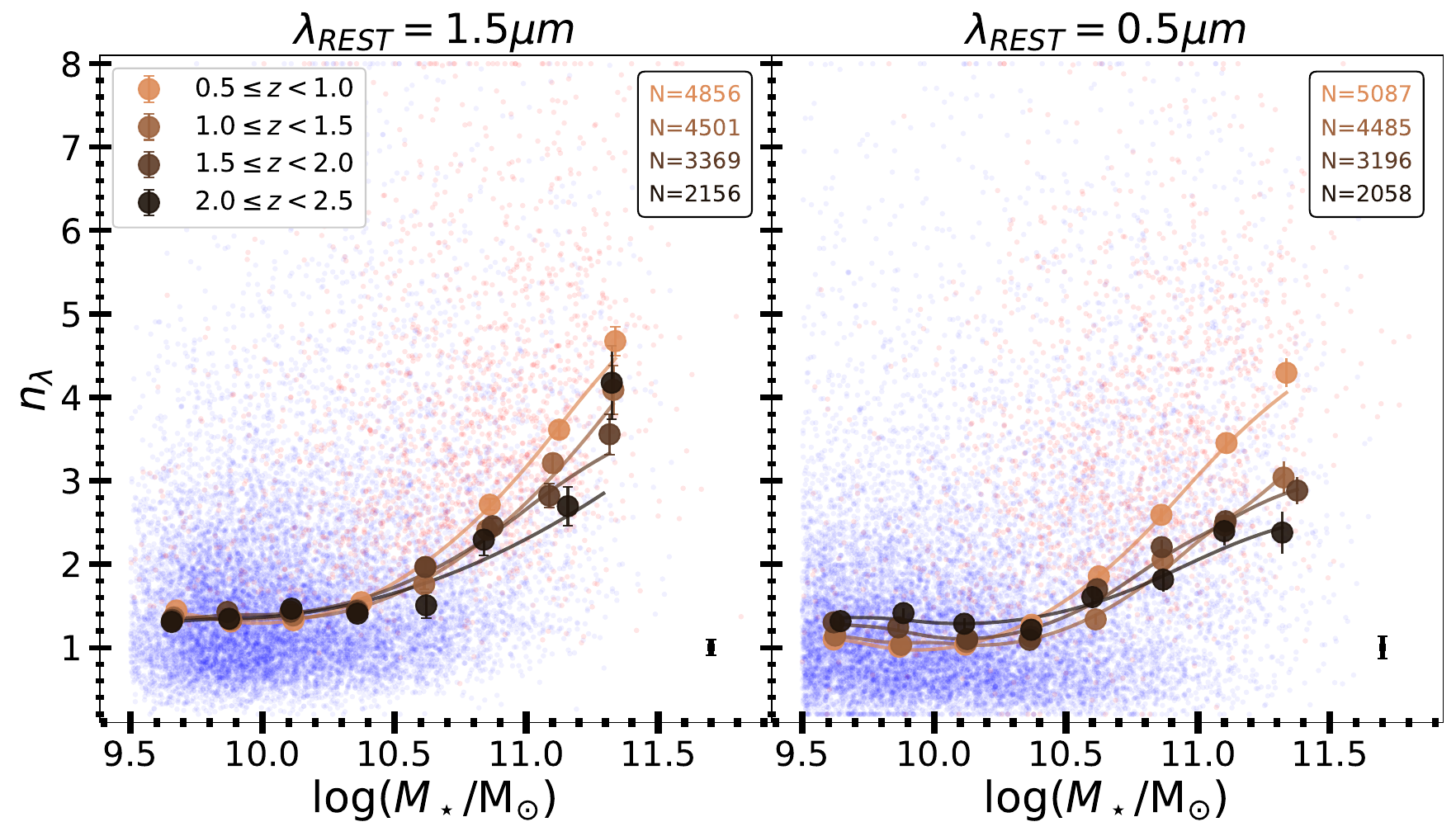}
        \caption{S\'ersic index at rest-frame 1.5$\mu{\text{m}}$ (left panel) and rest-frame 0.5$\mu{\text{m}}$ (right panel) as a function of stellar mass. 
        Dots in the background represent $n_{\lambda}$ of the individual star-forming (blue) and quiescent (red) galaxies.
        Filled circles show the median $n_{\lambda}$ in stellar mass bins of width 0.25~dex and four redshift bins from low-z (light) to high-z (dark).
        Error bars identify the statistical uncertainties computed as $\sigma/\sqrt{N}$ with $N$ the number of galaxies within the bin and $\sigma$ the standard deviation of the distribution.
        Solid lines show spline-quantile regression obtained using the COBS library \citep{ng07,ng22}. 16-50-84 percentiles of $n$ at each redshift and mass bins are reported in Appendix \ref{appendix: tables}.
        The median uncertainty on the S\'ersic index is shown in the bottom right corner as a black errorbar. We report the number of galaxies in each redshift bin in the top right corner of each panel.
        In any redshift bin massive galaxies have higher $n_{\lambda}$ than at lower mass.}
    \label{fig:n-mass}
    \end{figure*}

    Figure \ref{fig:n-mass} presents the rest-frame near-IR and optical S\'ersic index ($n_{1.5\mu{\text{m}}}$ and $n_{0.5\mu{\text{m}}}$, respectively) dependence on stellar mass in four redshift bins, from $z=0.5$ to $z=2.5$. 
    When considering the full galaxy population (combining star-forming and quiescent galaxies) below $M_\star=10^{10.3}~{\text{M}}_\odot$, both $n_{0.5\mu{\text{m}}}$ and $n_{1.5\mu{\text{m}}}$ have a median $n\approx 1.4$ in all redshift bins. At higher masses, $n$ steadily increases, up to $\approx 3-5$ at $M_\star=10^{11.3}~{\text{M}}_\odot$.
    As highlighted by the dots in the figure's background, part of this trend with mass is associated with the increased fraction of quiescent galaxies at high mass \citep[e.g.,][]{bundy06, ilbert10, muzzin13}, which have systematically higher S\'ersic indices compared to star-forming galaxies \citep{blanton03, franx08,wuyts11,bell12,barro17,whitaker17,martorano23}. 
    We will further address the different behavior of star-forming and quiescent galaxies, and the relation between star-formation activity and structure, in section \ref{sec:n-SFR}.

    $n_{1.5\mu{\text{m}}}$ and $n_{0.5\mu{\text{m}}}$ show very similar trends with stellar mass and redshift \citep[also see][]{martorano23}. Only at high mass and at $z>1$ there is a mild wavelength dependence, with  $n_{1.5\mu{\text{m}}} > n_{0.5\mu{\text{m}}}$. In the highest-mass bin ($M_\star>10^{11.3}~{\text{M}}_\odot$) the median near-IR S\'ersic index reaches $n_{1.5\mu{\text{m}}}\approx 4$, similar to $z<1$, while in the optical the mass dependence is flatter, reaching  $n_{0.5\mu{\text{m}}}\approx 3$, distinctly lower than at $z<1$.
    
\begin{figure*}
        \centering
        \includegraphics[scale=0.5]{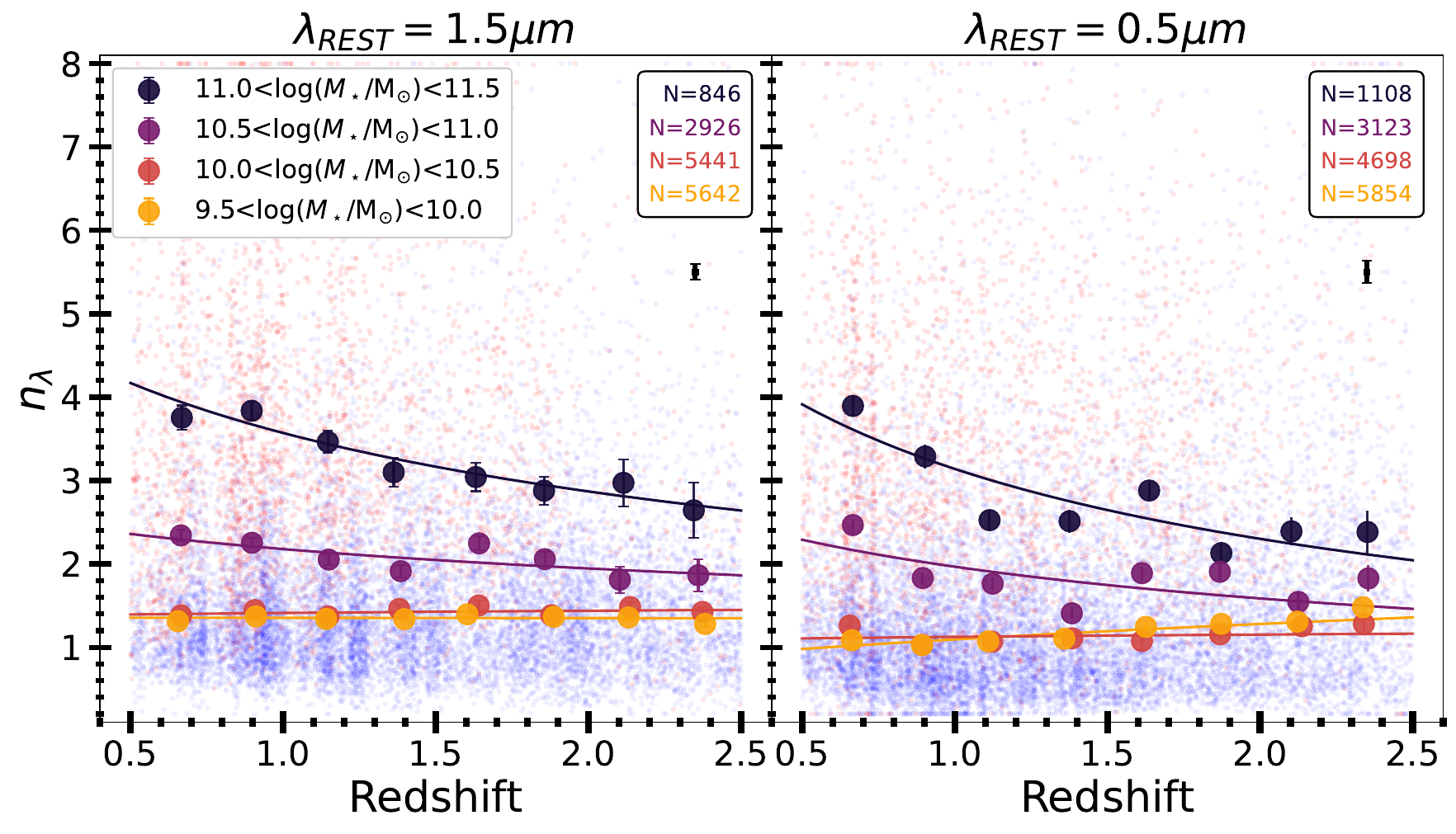}
        \caption{S\'ersic index at rest-frame 1.5$\mu{\text{m}}$ (left panel) and rest-frame 0.5$\mu{\text{m}}$ (left panel) as a function of redshift. 
        Dots in the background represent $n_{\lambda}$ of the individual star-forming (blue) and quiescent (red) galaxies.
        Filled circles show the median $n_{\lambda}$ in redshift bins of width 0.25 and in four stellar mass bins from high-M$_\star$ (dark) to high-M$_\star$ (light).
        Error bars identify the statistical uncertainties computed as $\sigma/\sqrt{N}$ with $N$ the number of galaxies within the bin and $\sigma$ the standard deviation of the distribution.
        Solid lines shows results of the fits to $n_\lambda\propto(1+z)^{\beta_\lambda}$.
        In the top right corner, we report the number of galaxies in each stellar-mass bin using the same color coding, and we show the median uncertainty on the S\'ersic index as a black errorbar.
        The S\'ersic index of massive galaxies evolves with redshift while that of lower mass galaxies does not.}
    \label{fig:n-z}
    \end{figure*}

    Put differently, at high mass, $n_{0.5\mu{\text{m}}}$ evolves somewhat more strongly with redshift than $n_{1.5\mu{\text{m}}}$.
    Figure \ref{fig:n-z} visualizes the same data as Figure \ref{fig:n-mass}, but now shown as a function of redshift in bins of stellar mass. To quantify the evolution of $n$ with redshift, we parameterize the S\'ersic index evolution across cosmic time as $n_\lambda\propto(1+z)^{\beta_\lambda}$.
    For each stellar mass bin, we fit $1\,000$ times the median S\'ersic index computed in redshift bins of width randomly sampled in the range 0.05-0.5, and weighing each value by the inverse of the statistical uncertainty. The best-fit coefficient (first two columns of Table \ref{tab:fit_res}) is given by the median of the parameters obtained in the $1\,000$ iterations of the fit.
    The retrieved values for $\beta_\lambda$ confirm a statistically stronger redshift evolution in the optical than in the near-IR for galaxies with $M_\star>10^{10.5}~{\text{M}_\odot}$ and a negligible evolution ($\beta_\lambda\simeq0$) for galaxies with $10^{10}<M_\star/{\text{M}_\odot}<10^{10.5}$.

    The absence of any redshift evolution at $M_\star<10^{10.5}~{\text{M}}_\odot$ is striking, especially considering that the intrinsic 3D shapes evolve quite radically over this redshift range \citep{van-der-wel14a, zhang19, pandya24}, with predominantly flattened disks at $z<1$ and a common occurrence of elongated (prolate) shapes at $z>1.5$. Despite this fundamental change in shape, the radial profile remains approximately exponential. It is important to keep in mind that, especially at $z>1$, $n\approx 1$ does not (necessarily) imply a disk-like morphology/geometry: the general correspondence between exponential profiles and diskiness applies to galaxies in this mass range and in the present-day Universe, but not generally.

    The mass dependence that exists regardless of redshift, combined with the notion that galaxies grow through star formation and/or merging, implies that individual galaxies increase their $n$ over time, even if, as is the case, the correlation between $n$ and stellar mass shows little redshift dependence.

    The lack of (or mild) redshift evolution of $n_{1.5\mu{\text{m}}}$, at a fixed stellar mass, suggests the different conditions at earlier cosmic times do not play a dominant role in defining the radial profiles of galaxies. In fact, even at much earlier cosmic times galaxies (up to $z\sim10$) have approximately exponential profiles \citep{robertson23, morishita24}. Instead, at all cosmic times the (radial) structure is related to mass and star-formation activity, as we will discuss below in Section \ref{sec:n-SFR}.

\subsection{The relation between S\'ersic index and SFR}{\label{sec:n-SFR}}
    \begin{figure*}
        \centering
	\includegraphics[scale=0.4]{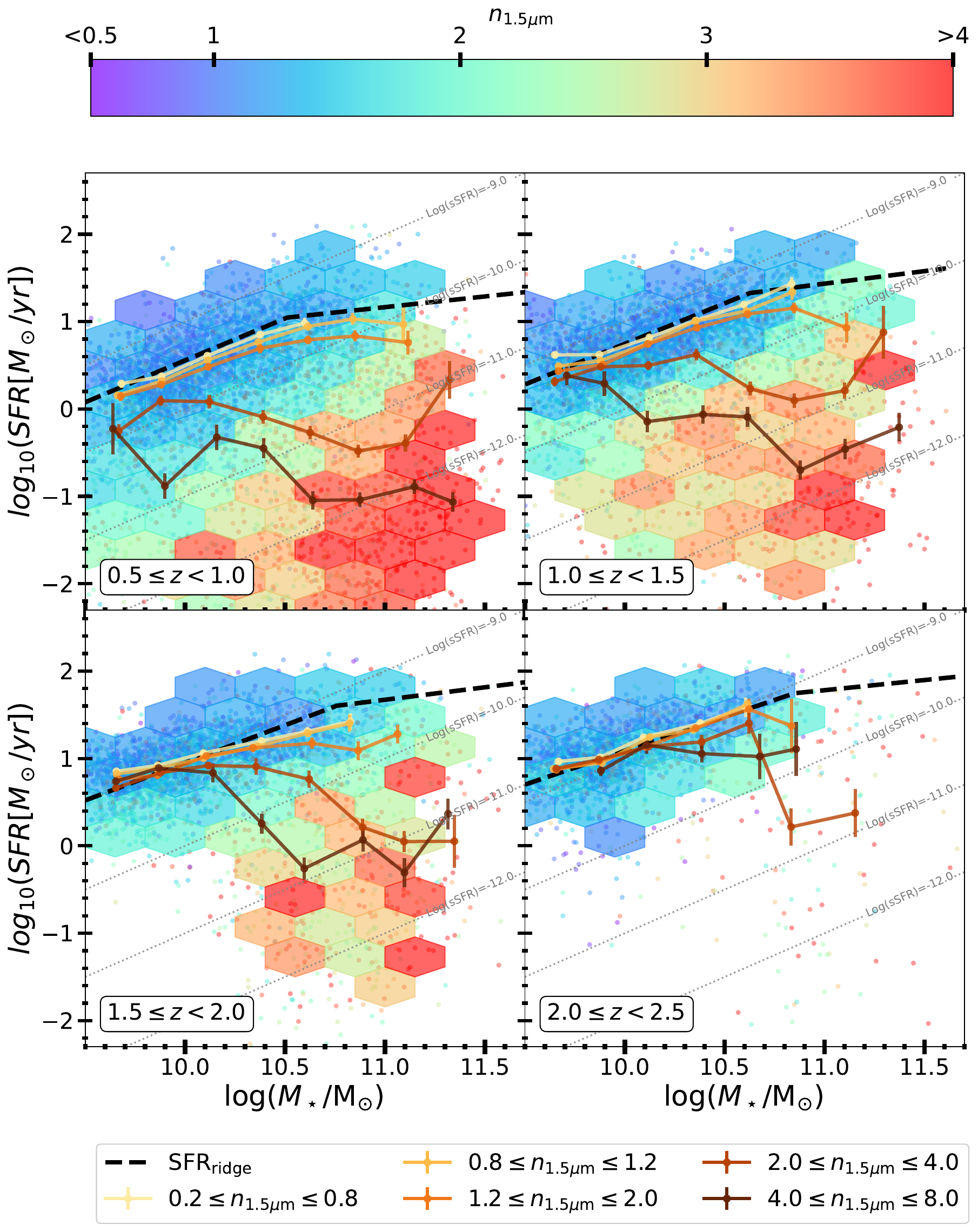}
        \caption{Star-formation rate (SFR) vs stellar mass in four redshift bins color-coded by $n_{1.5\mu{\text{m}}}$. The dashed black line represents the SFR-ridge identified in \cite{leja20}. Solid lines show the median trends in five $n_{1.5\mu{\text{m}}}$ bins. Hexbins are drawn around groups of at least 10 galaxies and colored with the median $n_{1.5\mu{\text{m}}}$. For reference, constant $\log({\text{sSFR}})$ lines are shown in light grey.
        Exponential-like galaxies lay on the SFMS at any redshift and stellar mass. $n_{1.5\mu{\text{m}}}>2$ galaxies detach from the SFMS at different stellar masses as a function of redshift.
        }
    \label{fig:SFR_M_ccN}
    \end{figure*}

    In the previous section, we showed how the optical and near-IR S\'ersic index depends on stellar mass and redshift, without considering star-formation activity, which is generally known to play a key role in understanding galaxy structure, both for present-day galaxies \citep{kauffmann03a} and at earlier cosmic times \citep{franx08, wuyts11, bell12, whitaker17}. In Figures \ref{fig:n-mass} and \ref{fig:n-z} we already preempted the connection with star-formation activity: quiescent galaxies clearly have higher $n$ values than star-forming galaxies, regardless of mass and redshift. Before presenting a split analysis of the mass and redshift dependence for quiescent and star-forming galaxies separately, we first analyze the correlation with SFR.
    
    Figure \ref{fig:SFR_M_ccN} shows the star-formation rate - stellar mass diagram, color-coded by near-IR S\'ersic index.
    At all redshifts and for all stellar masses there is a correspondence between star-formation rate and median $n_{1.5\mu{\text{m}}}$: galaxies with lower star formation typically have higher $n_{1.5\mu{\text{m}}}$. Likewise, as indicated by the running medians in Fig \ref{fig:SFR_M_ccN} (colored lines), galaxies with higher $n_{1.5\mu{\text{m}}}$ have lower SFR at the high-mass end. At low mass ($M_\star\lesssim 10^{10}~{\text{M}_\odot}$) and $z>1$ this stratification disappears: regardless of $n_{1.5\mu{\text{m}}}$ the median SFR is the same. But note that the trend between $n_{1.5\mu{\text{m}}}$ and SFR is still present. This implies that the bulk of these low-mass galaxies is on the star-forming sequence, but that among high-$n_{1.5\mu{\text{m}}}$ galaxies a tail toward low SFR also exists that is absent among low-$n_{1.5\mu{\text{m}}}$ galaxies. In any case, high-$n_{1.5\mu{\text{m}}}$, low mass galaxies are rare (see Fig.~\ref{fig:n-mass}).
    
    The figure further shows that lines of constant specific SFR  (shown in Figure \ref{fig:SFR_M_ccN} as dotted light-grey lines), down to $\log(\text{sSFR})\sim-10$, have an approximately constant $n_{1.5\mu{\text{m}}}$.
    The increase of $n_{1.5\mu{\text{m}}}$ with mass for galaxies on the star-forming sequence (or ridge) is associated with the bending of the sequence/ridge, not with a correlation between structure and mass at fixed sSFR. This is reminiscent of the result that disk components of galaxies have a linear star-forming sequence, at least in the local Universe \citep{nair11}.

    \begin{figure*}
        \centering
	\includegraphics[scale=0.33]{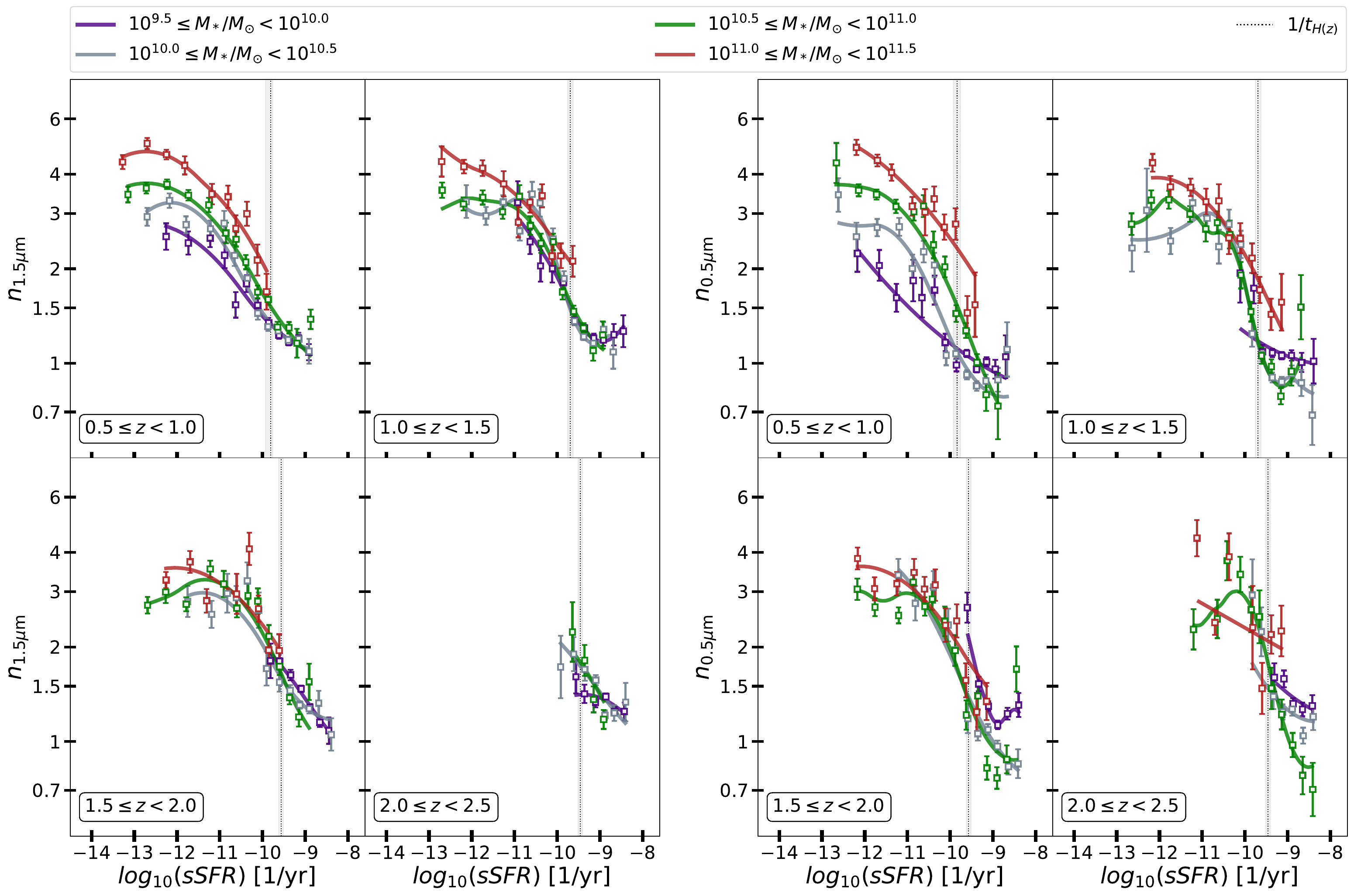}
        \caption{S\'ersic index as a function of the specific Star Formation Rate (sSFR) in four redshift bins. Each color represents a stellar mass bin with solid lines showing spline-quantile regression and squares showing the median S\'ersic index in sSFR bins. Error bars show the statistical uncertainty on the median ($\sigma/\sqrt{N}$). We highlight the $\frac{1}{t_H(z)}$ values between the two redshift extremes as a grey-shaded area and as a dotted line the position at the median redshift. Systematic differences in $n_{0.5\mu{\text{m}}}$ (left) and $n_{1.5\mu{\text{m}}}$ (right) due to stellar mass appear at $z\leq1$ independently on the wavelength observed.}
    \label{fig:n-Delta(SFR)}
    \end{figure*}

    The dependence (or lack thereof) of $n_{1.5\mu{\text{m}}}$ on stellar mass at fixed sSFR is made explicit in Figure \ref{fig:n-Delta(SFR)}. The dominant trend is that lower sSFR correlates with higher $n$~values, but a striking new feature appears: while at $z<1$ there is a strong mass dependence in the $n$~values at a fixed sSFR, at $z>1$ there is not. At $z>1$, $n$ only depends on sSFR, and not on stellar mass. 
    
    Figure \ref{fig:n-Delta(SFR)} also elucidates the connection between star-formation activity and the emergence of peaked light profiles (bulge-dominated systems).
    At any redshift, there is an anti-correlation between $n$ and sSFR, namely that star formation activity is suppressed in galaxies with more centrally concentrated light profiles. This supports the general picture in which star formation declines for galaxies with prominent bulges and/or a spheroidal structure \citep{huertas-company16}.
    The trend for $n_{1.5\mu{\text{m}}}$ in the redshift bin $1.5\leq z<2$ is particularly striking: over more than four orders of magnitude, the sSFR tightly correlates with $n_{1.5\mu{\text{m}}}$ with no discernible dependence on mass. The increase of $n$ with mass seen in Figure \ref{fig:n-mass} is, therefore, driven by the underlying anti-correlations $n$-sSFR and mass-sSFR.
    
    At later cosmic times ($z < 1$) there is a slight dependence on stellar mass, in the sense that, at fixed sSFR, higher-mass galaxies have larger $n$. This trend is more pronounced for $n_{0.5\mu{\text{m}}}$ than for $n_{1.5\mu{\text{m}}}$, suggesting that $M_\star/L$ gradients play a role that potentially extends to the near-IR. 

    \begin{figure*}
        \centering
	\includegraphics[scale=0.5]{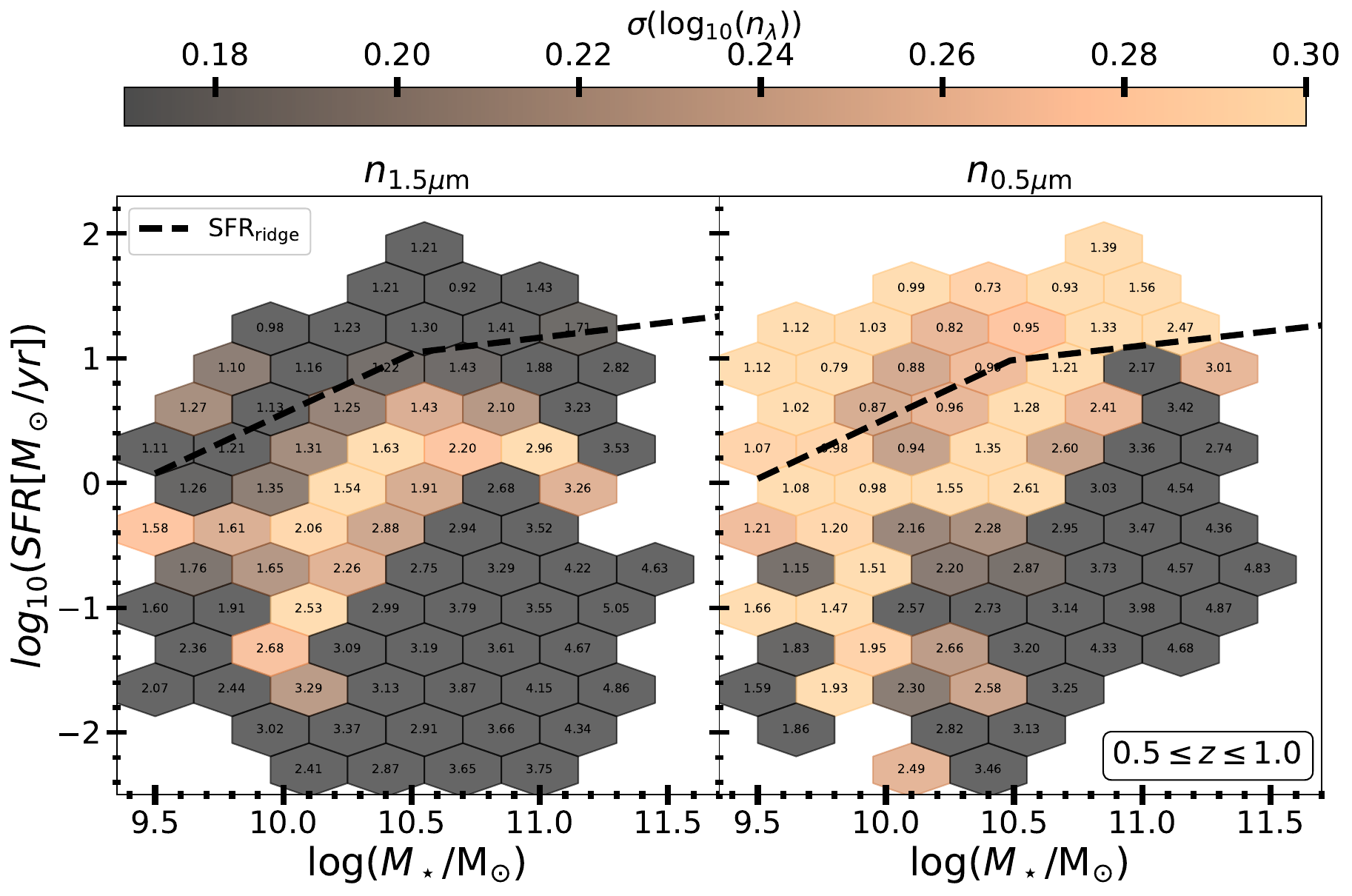}
        \caption{SFR-stellar mass plane for galaxies at redshifts $0.5\leq z<1$ in hexbins, color-coded by half of the 16-84th percentile range of $\log_{10}(n_\lambda)$ computed over at least 10 galaxies. Numerical values represent the median $n_\lambda$ of galaxies within the hexbin. \emph{Left:} $n_{0.5\mu{\text{m}}}$; \emph{Right:}  $n_{1.5\mu{\text{m}}}$. $n_{1.5\mu{\text{m}}}$ only shows a large scatter in the green valley while $n_{0.5\mu{\text{m}}}$ has a large scatter also for galaxies on the SFMS.}
    \label{fig:n_dispersion}
    \end{figure*}

\subsubsection{Scatter in S\'ersic indices}
    The results described above refer to median values of $n$, but there is considerable spread in $n$ values at fixed redshift, mass, and star-formation rate. The scatter in $\log n$ across the SFR-stellar mass plane is shown in Figure \ref{fig:n_dispersion}. For $n_{1.5\mu{\text{m}}}$ the scatter around the median values is $\approx 0.18$ dex for galaxies on the SFMS as well as those far ($>$1~dex) below it, but significantly elevated to $\approx 0.25$~dex in the green valley, the region $0.5-0.7$~dex below the SFMS. The variety in structural properties peaks in this transitionary region, implying a true, physical variation in structure as galaxies follow a variety in pathways toward quiescence \citep{wu18b}, perhaps with a contribution from galaxies that are undergoing rejuvenation events, that is periods of renewed, elevated, star formation activity after a quiescent phase \citep[e.g.,][]{chauke19, mancini19}.

    The scatter in $n_{0.5\mu{\text{m}}}$ for galaxies on the star-forming sequence is much larger than the scatter in $n_{1.5\mu{\text{m}}}$ ($\approx0.27$~dex vs.~0.18 dex): differing viewing angles and dust properties, perhaps combined with a larger variety of stellar populations properties, result in extra scatter in the observed light profiles \citep[see e.g.][]{zhang23} on top of underlying variations in the stellar mass profiles. This trend obfuscates the increased scatter in galaxy structure in the green valley. 
    
    At $z>1$ the peak in scatter in the green valley is less clear (not shown here), and it remains to be seen whether this is physical or the result of limitations in the data due to the smaller sample sizes in that region of the stellar mass-SFR plane and the larger measurement uncertainties.

\subsubsection{Separating quiescent and star-forming galaxies}

    \begin{figure*}
        \centering
	\includegraphics[scale=0.5]{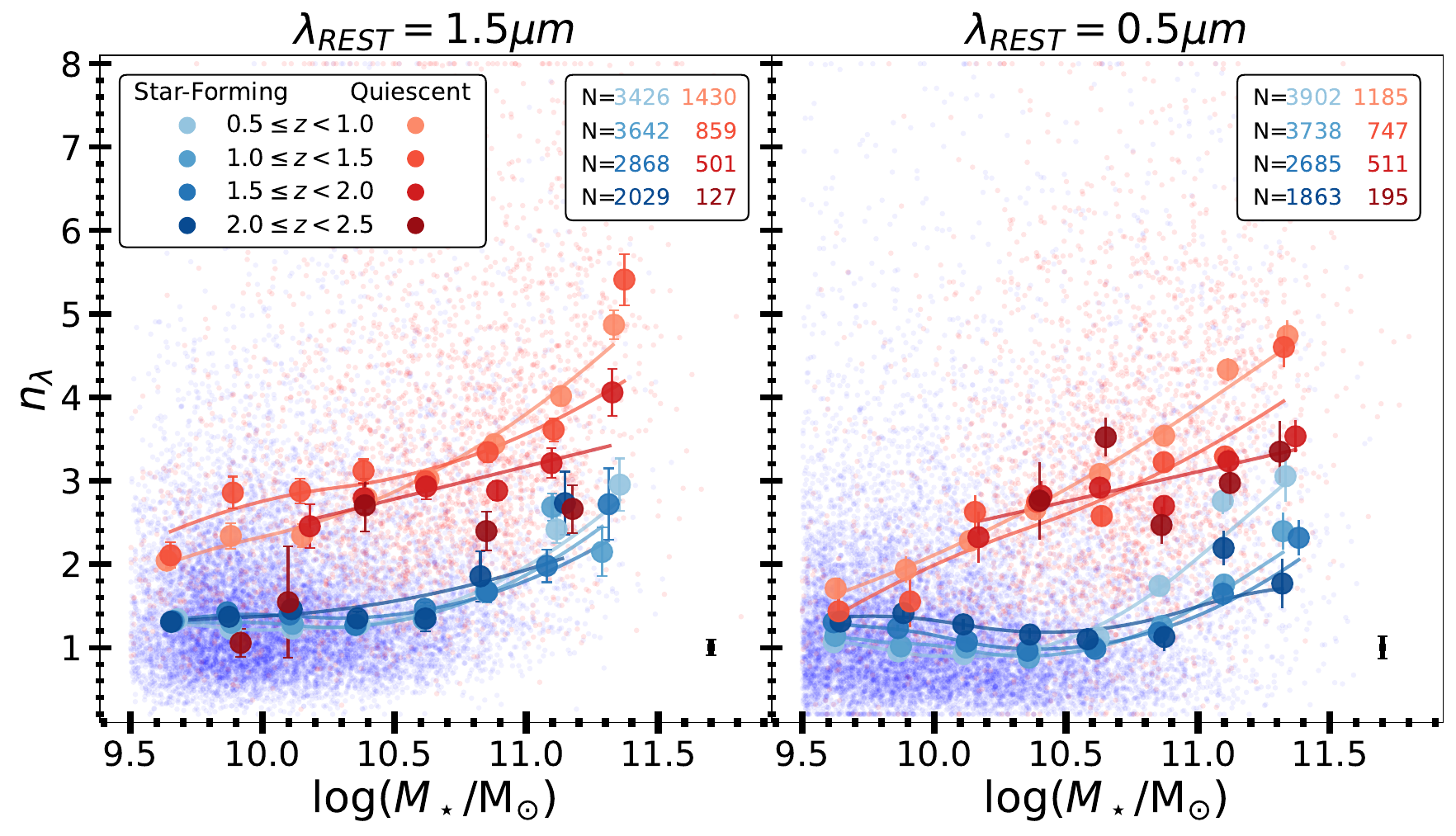}
        \caption{Same as Figure \ref{fig:n-mass} but medians are now computed separately on the quiescent (reds) and star-forming (blues) samples. Solid lines show spline-quantile regression obtained using the COBS library \citep{ng07,ng22}. 16-50-84 percentiles of $n$ at each redshift and mass bins are reported in appendix \ref{appendix: tables}. For the quiescent population in the highest redshift bin, just the circles are shown and not the solid line because the spline regression was not robust enough given the low number of samples.
        The median uncertainty on the S\'ersic index is shown in the bottom right corner as a black errorbar. We report the number of galaxies in each redshift bin in the top right corner of each panel.
        $n_{\lambda}$ evolves with stellar mass for both populations.}
    \label{fig:n-massSFQ}
    \end{figure*}

    \begin{figure*}
        \centering
	\includegraphics[scale=0.5]{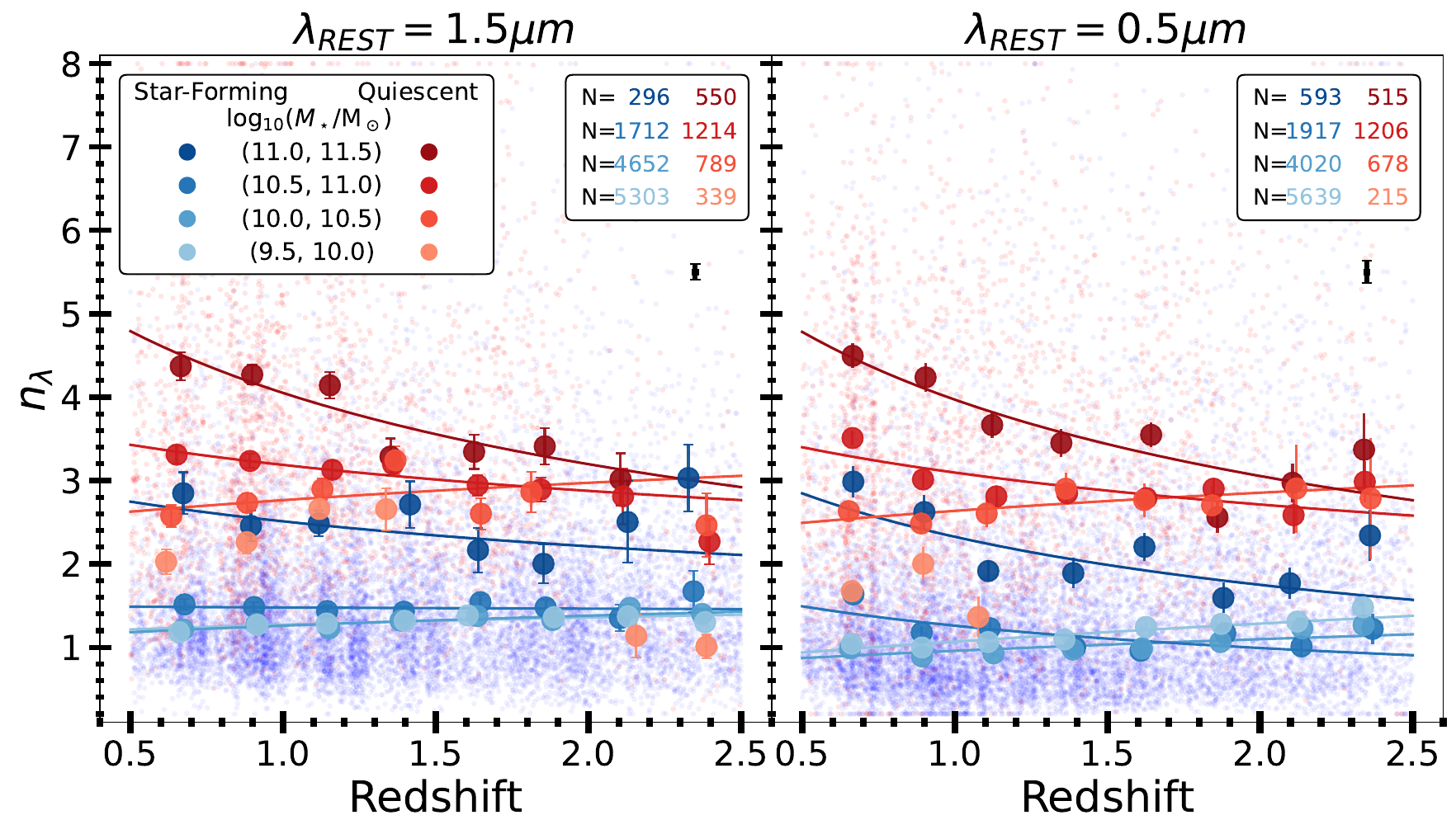}
        \caption{Same as Figure \ref{fig:n-z} but medians are now computed separately on the quiescent (reds) and star-forming (blues) samples. Solid lines represent fits to $n_\lambda\propto(1+z)^{\beta_\lambda}$. Fits to the lowest-mass bin for quiescent galaxies are not shown because considered not robust due to the small sample. 
        In the top right corner, we report the number of galaxies in each stellar-mass bin using the same color coding, and we show the median uncertainty on the S\'ersic index as a black errorbar.
        For both populations, $n_{\lambda}$ mildly evolves with redshift for low-M$_\star$ galaxies while it does for the ones at high-M$_\star$.}
    \label{fig:n-zSFQ}
    \end{figure*}

    Now that the relation between SFR and S\'ersic index has been explored, we examine the mass and redshift dependence of $n$ for quiescent and star-forming galaxies separately.  Figure \ref{fig:n-massSFQ} shows that, as anticipated, for any mass and redshift interval, quiescent galaxies have, on average, higher $n_{1.5\mu{\text{m}}}$ than star-forming galaxies. Both populations show an increase in $n_{1.5\mu{\text{m}}}$ with increasing stellar mass. For quiescent galaxies this increase is monotonic, while for star-forming galaxies the relation is flat up to $M_\star\approx 10^{10.5}~{\text{M}}_\odot$ and increases at higher mass. 

    We note that the increase in $n_{1.5\mu{\text{m}}}$ with $M_\star$ at $z>1$ (already seen in Fig.~\ref{fig:n-mass}) for star-forming galaxies does not contradict the lack of a stellar mass-dependence in the $n$-sSFR plane (Fig.~\ref{fig:n-Delta(SFR)}): the distinction between star-forming and quiescent galaxies is based on offset from the SFMS, which has a sub-linear slope at high mass, adding lower-sSFR (that is higher-$n$) to the population of star-forming galaxies at high stellar mass.

    The results presented in Figure \ref{fig:n-massSFQ} reproduce previous findings based on UV-optical profile analysis \citep[e.g.,][]{wuyts11,bell12,lang14, barro17, whitaker17} that at all redshifts $z\leq 3$ quiescent galaxies, on average, have more peaked light profiles than star-forming galaxies, and that high-mass galaxies have higher $n$ than lower-mass galaxies.
    As previously shown by \cite{martorano23}, intrinsic differences up to $50\%$ between $n_{1.5\mu{\text{m}}}$ and $n_{0.5\mu{\text{m}}}$ are present, but the distinction in the radial light profile between quiescent and star-forming galaxies holds in the near-IR as well.
    The persistence of a structural difference between star-forming and quiescent galaxies in the rest-frame near-IR implies a true, physical difference in structure: if the structural difference seen in the rest-frame optical were merely apparent due to the presence of bright, exponential disks in star-forming galaxies that fade upon the cessation of star formation, then the structural difference would be less apparent in the near-IR and in mass-weighted profiles \citep[see also][]{martorano23, van-der-wel24}.

    For massive star-forming galaxies at $z>1$, $n_{1.5\mu{\text{m}}}$ is systematically larger than $n_{0.5\mu{\text{m}}}$. These are also the galaxies whose sizes show the strongest wavelength dependence \citep{van-der-wel24, martorano24} and which are often seen to have strongly attenuated centers \citep{nelson16, miller22, lebail23}. This evidence, taken together, points at radially varying dust attenuation as an important factor for the structure measured in the rest-frame optical for massive, star-forming galaxies \citep[see also e.g.][]{zhang23, nelson23}. This is also supported by the analysis presented in \cite{nedkova24a} of the rest-frame UV and optical S\'ersic profiles of CANDELS galaxies, where authors find that the presence of centrally concentrated dust in massive galaxies flattens the light profile leading to a lower UV S\'ersic index than in the optical.
    At $z<1$ the median optical and near-IR S\'ersic indices for these massive star-forming galaxies are comparable, suggesting lower optical depths; galaxy sizes still differ substantially between the optical and near-IR \citep{van-der-wel24}. The implied $M_\star/L$ gradient is likely partially explained by stellar age/metallicity gradients, which is reproduced by radiative transfer calculations of simulated galaxies \citep{baes24b}.

    \begin{table*}
         \caption{$\beta_\lambda$ values from the parametrization $n_\lambda\propto(1+z)^{\beta_\lambda}$.}
        \label{tab:fit_res}
        \centering
        \begin{tabular}{|c|c|c|c|c|c|c|}
             \hline
             &\multicolumn{2}{c|}{FULL-SAMPLE} & \multicolumn{2}{c|}{STAR-FORMING} & \multicolumn{2}{c|}{QUIESCENT} \\
             \hline
             Mass bin& $\beta_{0.5\mu{\text{m}}}$ & $\beta_{1.5\mu{\text{m}}}$ & $\beta_{0.5\mu{\text{m}}}$ & $\beta_{1.5\mu{\text{m}}}$ & $\beta_{0.5\mu{\text{m}}}$ & $\beta_{1.5\mu{\text{m}}}$ \\
             \hline
             (9.5-10)& $ 0.39^{+0.04}_{-0.04}$&$ 0.00^{+0.04}_{-0.04}$&$ 0.45^{+0.04}_{-0.04}$&$ 0.16^{+0.05}_{-0.04}$&--&--\\ [1.5ex]
             (10-10.5)&$ 0.06^{+0.05}_{-0.05}$&$ 0.05^{+0.04}_{-0.04}$&$ 0.34^{+0.05}_{-0.05}$&$ 0.22^{+0.04}_{-0.04}$&$ 0.19^{+0.13}_{-0.13}$&$ 0.18^{+0.13}_{-0.13}$\\ [1.5ex]
             (10.5-11)&$-0.53^{+0.08}_{-0.08}$&$-0.27^{+0.07}_{-0.07}$&$-0.59^{+0.10}_{-0.11}$&$-0.02^{+0.07}_{-0.07}$&$-0.33^{+0.08}_{-0.08}$&$-0.25^{+0.08}_{-0.08}$\\ [1.5ex]
             (11-11.5)&$-0.77^{+0.10}_{-0.10}$&$-0.54^{+0.10}_{-0.10}$&$-0.71^{+0.16}_{-0.18}$&$-0.30^{+0.20}_{-0.23}$&$-0.65^{+0.10}_{-0.10}$&$-0.59^{+0.12}_{-0.11}$\\ [1.5ex]
            \hline

        \end{tabular}
        \tablefoot{
        Values of $\beta_\lambda$ represent the median and 16-84 percentile intervals retrieved as outlined in Sect. \ref{sec:n-m-z} for the Full-Sample (first two columns), the star-forming sample (middle two columns) and the quiescent sample (last two columns) in four stellar mass bins.
        }
    \end{table*} 
    
    Figure \ref{fig:n-zSFQ}, like Figure \ref{fig:n-z}, shows the redshift dependence of the S\'ersic index, but now dividing the sample into star-forming and quiescent galaxies. 
    As done for Figure \ref{fig:n-z}, we parameterize the S\'ersic index evolution across cosmic time as $n_\lambda\propto(1+z)^{\beta_\lambda}$ (results are shown in Table \ref{tab:fit_res}).
    We note that these findings do not depend on the definition of quiescence; all fitting results for $\beta_\lambda$ are not affected by more than 1$\sigma$ if we use other quiescence criteria based on the galaxy's sSFR (i.e. defining a galaxy as quiescent when $\log_{10}(\text{sSFR})<-11$ or $\log_{10}(\text{sSFR})<\frac{1}{3t_H(z)}$ with $t_H$ the age of the universe at the galaxy's redshift).
    
    Low-mass ($M_\star\lesssim 10^{10}~{\text{M}}_\odot$) star-forming galaxies show a subtle but interesting trend. The approximate result is that both optical and near-IR S\'ersic indices are typically approximately exponential at all redshifts, but there is a small but significant difference between $n_{1.5\mu{\text{m}}}$ and $n_{0.5\mu{\text{m}}}$ at $z<1$ that is absent at $z>1$. $n_{0.5\mu{\text{m}}}$ is smaller than $n_{1.5\mu{\text{m}}}$, which can either be explained by a significant concentration of dust in the center of these low-mass galaxies due to their high dust-formation and low dust-destruction efficiency \citep{calura16}, or age gradients due to young, star-forming outer parts. 
    
    At $M_\star>10^{10.5}~{\text{M}_\odot}$ the situation is more complicated, and the S\'ersic index depends on redshift, galaxy type, and wavelength. The S\'ersic index of star-forming galaxies shows a significant decrease with redshift, more so in the optical than in the near-IR. We attribute this to the evolution in centrally concentrated dust attenuation. 
    Conversely, for quiescent galaxies differences between $n_{0.5\mu{\text{m}}}$ and $n_{1.5\mu{\text{m}}}$ are minor, which is consistent with the notion that these objects are relatively poor in young stars and dust content.
    Furthermore, the fact that high-mass quiescent galaxies have somewhat lower $n$ values at high $z$ than at low $z$, is consistent with the idea that massive ellipticals gradually build up their outer parts through (dissipationless) merging \citep[e.g.,][]{naab09a, van-der-wel09, van-dokkum10c}.

    \cite{patel13} showed the redshift evolution of the rest-frame UV/optical S\'ersic index of star-forming and quiescent galaxies in the stellar mass range $10^{10.5}<M_\star/{\text{M}_\odot}<10^{12}$ in the redshift range 0.25-3. They found $n \propto (1+z)^\beta$ with $\beta=-0.50\pm0.18$ and $-0.64\pm0.16$ for quiescent and star-forming galaxies, respectively.     Limiting our sample to the same stellar mass range, in the redshift range $z=0.5-2.5$, we recover trends marginally compatible on the $1\sigma$ level, with $n_{0.5\mu{\text{m}}}\propto(1+z)^{-0.40\pm0.06}$ and $\propto(1+z)^{-0.44\pm0.09}$. 
    Differences are most likely driven by a combination of factors: their smaller sample; the different SED fitting code used to derive stellar population parameters, the different pipeline used to retrieve S\'ersic indices, and the different definition of quiescence adopted. Despite all these differences, the general trends recovered are compatible within the uncertainties. For comparison, for the same sample we find $n_{1.5\mu{\text{m}}}\propto(1+z)^{-0.34\pm0.07}$ and $\propto(1+z)^{-0.04\pm0.07}$ for quiescent and star-forming galaxies, respectively.

\section{Summary and conclusion}\label{sec: conclusion}

    We measure the rest-frame near-IR ($1.5\mu{\text{m}}$) S\'ersic indices of $\sim15\,000$ galaxies in the redshift range $0.5\leq z\leq2.5$, selected from the COSMOS-Web and PRIMER-COSMOS surveys with JWST/NIRCam (Section 2). The dependence on redshift, stellar mass, and star-formation activity of $n_{1.5\mu\text{m}}$ is compared with the rest-frame optical ($0.5\mu{\text{m}}$) from HST/CANDELS.  At fixed stellar mass up to $M_\star = 10^{10.5}~{\text{M}}_\odot$ the median S\'ersic index evolves slowly or not at all with redshift both in the optical and near-IR (Section 3.1). At higher masses ($M_\star>10^{11}~M_\odot$), both $n_{1.5\mu\text{m}}$ and $n_{0.5\mu\text{m}}$ evolve with redshift from $n\approx 2.5$ at $z=2.5$ to $n\approx 4$ at $z<1$. High-mass galaxies have higher $n$ than lower-mass galaxies (the sample reaches down to $M_\star=10^{9.5}~M_\odot$) at all redshifts, with a stronger dependence in the rest-frame near-IR than in the rest-frame optical at $z>1$ (Section 3.1). This wavelength dependence is caused by star-forming galaxies that, at $z>1$ but not $z<1$, have lower optical than near-IR $n$. Star-forming galaxies generally have lower $n$ than quiescent galaxies, also in the near-IR, confirming and fortifying the result that there exists a connection between star-formation activity and radial stellar mass profile across cosmic time. Besides these general trends that confirm previous results, two new trends emerge: 1) at $z>1$ the median near-IR $n$ varies strongly with star formation activity, but not with stellar mass (Section 3.2), and 2) the scatter in near-IR $n$ is substantially higher in the green valley (0.25 dex) than on the star-forming sequence and among quiescent galaxies (0.18 dex) -- this trend is not seen in the optical because dust and young stars contribute to the variety in optical light profiles (Section 3.2.1). 
 
    The variety of physical processes and evolutionary pathways of individual galaxies imply that a unique interpretation of the trends presented in this paper is not possible and conclusions based on observations such as those presented in this paper will always remain speculative. General tendencies may be identified but more insight must come from the comparison with simulations \citep[i.e.][]{martig09, ceverino10, wetzel13, tacchella19, pillepich19, gargiulo22, park22a, wang23, bluck23}. Much attention has been given to interpreting the size evolution of galaxies in the context of simulations \citep[and others]{furlong15, genel18, lee23, costantin23}. However, S\'ersic indices (or related parameters that quantify concentration) of simulated galaxies in the same redshift range investigated in this work, are not often shown or calculated. \citet{wuyts10} found that radial profiles of simulated galaxies compared poorly with observed profiles, but in the meantime, simulations have improved in resolution and treatment of physical processes. \citet{tacchella16} showed that the mass profiles of 26 simulated galaxies resemble those of observed galaxies across a range in redshifts, but at $z=2$ that sample contains no $M_\star > 10^{10.7} M_\odot$ galaxies, which is where most variation and evolution with redshift is seen in the observations.
    Now that near-IR radial profiles are available due to JWST/NIRCam observations, the next step must be to quantify the radial profiles of large samples of simulated galaxies with sufficient spatial resolution and compare with the newly measured (and tabulated) data in this paper.

\begin{acknowledgements}
      MM acknowledges the financial support of the Flemish Fund for Scientific Research (FWO-Vlaanderen), research project G030319N.
\end{acknowledgements}

\bibliographystyle{aa} 
\bibliography{mypapers.bib} 

\begin{appendix} 
    \onecolumn
    \section{Datasets comparison}\label{appendix: dataset_comp}
    In this work, we compare the S\'ersic index for galaxies retrieved from two different samples for which stellar parameters were retrieved by different \textsc{Prospector} runs using different photometric catalogs. In this appendix, we compare the stellar mass and SFR retrieved in this work and in \cite{leja20} for a subset of $1\,656$ galaxies that fall in the PRIMER-COSMOS field and for which values from both the \textsc{Prospector} runs are available.
    
    \begin{figure*}[h!]
        \centering
    	\includegraphics[scale=0.275]{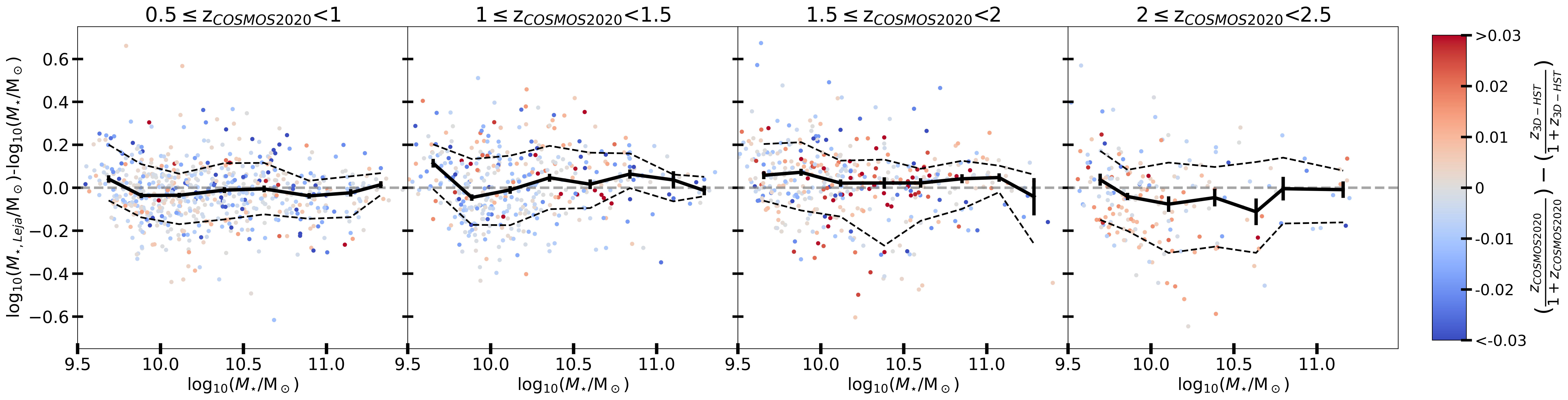}
            \caption{Difference in stellar mass measured in this work and in \cite{leja20} as a function of the stellar mass retrieved in this work in 4 redshift bins. Color coding conveys the difference of the relative redshift estimates from the COSMOS2020 catalog (used in this work) and in the 3D-HST catalog \citep[used in][]{leja20}. The solid black line shows the median difference with error bars representing the statistical uncertainty $\sigma/\sqrt{N}$ while dashed lines show the 16-84 percentile range. A systematic difference appears in the highest redshift bin mostly driven by a systematic $5\%$ difference in the redshift value adopted.}
        \label{fig:mass_comp}
        \end{figure*}
    
    \begin{figure*}[h!]
        \centering
    	\includegraphics[scale=0.275]{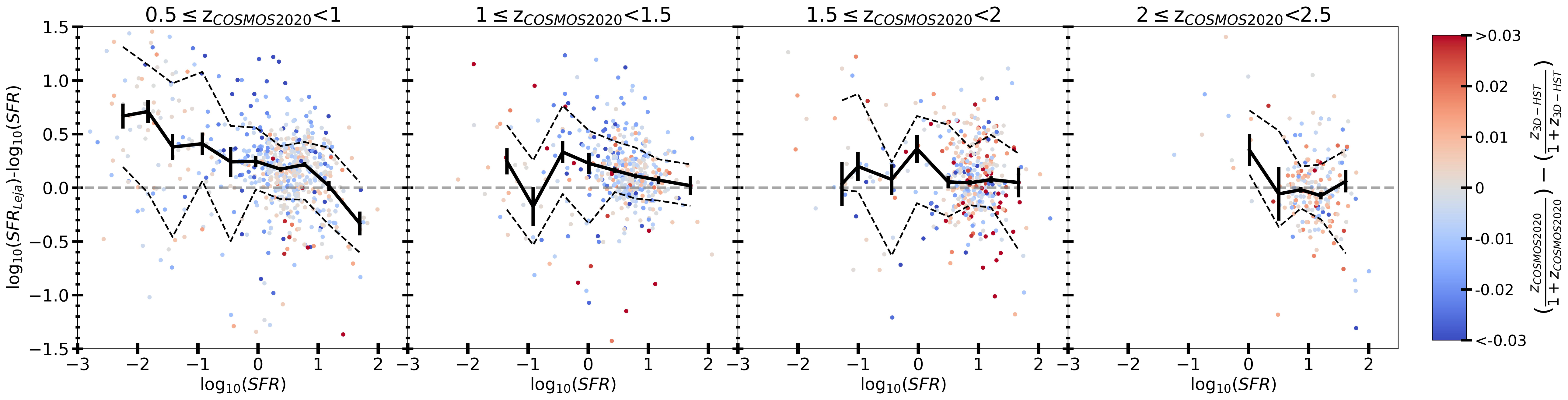}
            \caption{Difference in SFR measured in this work and in \cite{leja20} as a function of the SFR retrieved in this work in 4 redshift bins. Color coding conveys the difference of the relative redshift estimates from the COSMOS2020 catalog (used in this work) and in the 3D-HST catalog \citep[used in][]{leja20}. The solid black line shows the median difference with error bars representing the statistical uncertainty $\sigma/\sqrt{N}$ while dashed lines show the 16-84 percentile range. Systematic differences are of the order of 0.1-0.2~dex for high-SFR galaxies and up to 0.6~dex for low-SFR ones. These differences do not play a major role in this work.}
        \label{fig:sfr_comp}
        \end{figure*}
    
    As Figure \ref{fig:mass_comp} shows, systematics in stellar mass between the two catalogs are below 0.1~dex for $z<2$ and up to 0.1~dex for $z>2$ with this latter driven by a systematic $5\%$ difference in the redshift value adopted for the SED fit.
    
    Differences in the SFR estimates can be up to 0.6~dex (see Figure \ref{fig:sfr_comp}) for low-SFR galaxies and $\sim0.2$~dex for the high-SFR ones. These do not play a major role in this work.

    \section{Mass-redshift variation of S\'ersic index}\label{appendix: tables}
        In this appendix, we make available the medians and 16-84 percentiles of the lines drawn in Figures \ref{fig:n-mass} and \ref{fig:n-massSFQ}.
    
        \begin{table*}
    	\centering
    	\caption{This table quantifies lines presented in Figures \ref{fig:n-mass} and \ref{fig:n-massSFQ}}
    	\label{tab:med_nir}
            {\tiny
            \begin{tabular}{P{2cm}|P{0.75cm}P{0.75cm}P{1cm}|P{0.75cm}P{0.75cm}P{1cm}|P{0.75cm}P{0.75cm}P{1cm}|P{0.75cm}P{0.75cm}P{1cm}}
    		\hline
        	{} & \multicolumn{3}{c|}{$0.5\leq z<1$} & \multicolumn{3}{c|}{$1\leq z<1.5$} & \multicolumn{3}{c|}{$1.5\leq z<2$} & \multicolumn{3}{c}{$2\leq z<2.5$} \\
                \hline
                \rule{0pt}{0.3cm}
                Mass bin & ALL & SF & Q & ALL & SF & Q & ALL & SF & Q & ALL & SF & Q \\[0.4ex]
                \hline
                \rule{0pt}{0.4cm}
                $[10^{9.50}-10^{9.75}$) & 1.44$_{-0.59}^{+1.00}$ & 1.34$_{-0.55}^{+0.89}$ & 2.04$_{-0.77}^{+1.01}$ & 1.36$_{-0.59}^{+1.25}$ & 1.32$_{-0.57}^{+1.17}$ & 2.11$_{-0.69}^{+0.81}$ & 1.33$_{-0.58}^{+0.85}$ & 1.31$_{-0.57}^{+0.83}$ & -- & 1.31$_{-0.55}^{+0.84}$ & 1.31$_{-0.55}^{+0.84}$ & -- \\ [1.5ex]
                $[10^{9.75}-10^{10.00}$) & 1.31$_{-0.51}^{+0.98}$ & 1.22$_{-0.44}^{+0.83}$ & 2.34$_{-0.89}^{+1.98}$ & 1.33$_{-0.56}^{+1.03}$ & 1.30$_{-0.53}^{+0.95}$ & 2.86$_{-1.14}^{+1.47}$ & 1.43$_{-0.62}^{+0.94}$ & 1.42$_{-0.62}^{+0.94}$ & -- & 1.34$_{-0.59}^{+0.87}$ & 1.37$_{-0.61}^{+0.85}$ & 1.06$_{-0.44}^{+0.59}$ \\ [1.5ex]
                $[10^{10.00}-10^{10.25}$) & 1.33$_{-0.49}^{+1.15}$ & 1.23$_{-0.43}^{+0.82}$ & 2.34$_{-1.00}^{+2.07}$ & 1.38$_{-0.59}^{+1.31}$ & 1.28$_{-0.52}^{+1.00}$ & 2.87$_{-1.00}^{+2.00}$ & 1.43$_{-0.63}^{+1.04}$ & 1.40$_{-0.61}^{+1.03}$ & 2.45$_{-0.93}^{+1.96}$ & 1.47$_{-0.66}^{+0.97}$ & 1.46$_{-0.66}^{+0.97}$ & 1.54$_{-0.55}^{+2.50}$ \\ [1.5ex]
                $[10^{10.25}-10^{10.50}$) & 1.54$_{-0.63}^{+1.65}$ & 1.28$_{-0.46}^{+1.03}$ & 2.79$_{-1.08}^{+2.05}$ & 1.45$_{-0.58}^{+1.55}$ & 1.27$_{-0.47}^{+0.98}$ & 3.12$_{-0.99}^{+2.21}$ & 1.44$_{-0.62}^{+1.55}$ & 1.27$_{-0.51}^{+1.03}$ & 2.79$_{-0.92}^{+1.96}$ & 1.41$_{-0.68}^{+1.42}$ & 1.35$_{-0.63}^{+1.27}$ & 2.70$_{-1.73}^{+1.00}$ \\ [1.5ex]
                $[10^{10.50}-10^{10.75}$) & 1.97$_{-0.89}^{+1.83}$ & 1.39$_{-0.46}^{+1.19}$ & 3.02$_{-1.06}^{+1.81}$ & 1.76$_{-0.79}^{+1.77}$ & 1.34$_{-0.48}^{+1.13}$ & 2.99$_{-1.01}^{+2.17}$ & 1.96$_{-0.98}^{+1.87}$ & 1.46$_{-0.61}^{+1.23}$ & 2.93$_{-0.94}^{+2.36}$ & 1.51$_{-0.60}^{+1.85}$ & 1.35$_{-0.45}^{+1.46}$ & -- \\ [1.5ex]
                $[10^{10.75}-10^{11.00}$) & 2.71$_{-1.33}^{+1.67}$ & 1.66$_{-0.51}^{+1.39}$ & 3.44$_{-1.08}^{+1.81}$ & 2.41$_{-1.11}^{+1.75}$ & 1.67$_{-0.53}^{+1.17}$ & 3.34$_{-1.12}^{+1.44}$ & 2.46$_{-1.18}^{+1.73}$ & 1.67$_{-0.75}^{+1.63}$ & 2.88$_{-0.83}^{+1.71}$ & 2.29$_{-0.70}^{+1.76}$ & 1.85$_{-0.93}^{+2.22}$ & 2.40$_{-0.38}^{+1.35}$ \\ [1.5ex]
                $[10^{11.00}-10^{11.25}$) & 3.61$_{-1.66}^{+1.78}$ & 2.42$_{-1.09}^{+1.80}$ & 4.02$_{-1.41}^{+1.66}$ & 3.21$_{-1.25}^{+2.17}$ & 2.68$_{-1.16}^{+1.52}$ & 3.61$_{-1.05}^{+1.93}$ & 2.82$_{-1.20}^{+1.53}$ & 1.98$_{-0.70}^{+1.94}$ & 3.21$_{-0.94}^{+1.61}$ & 2.69$_{-0.93}^{+1.35}$ & 2.73$_{-1.20}^{+1.27}$ & 2.66$_{-0.66}^{+0.85}$ \\ [1.5ex]
                $[10^{11.25}-10^{11.50}$) & 4.67$_{-1.75}^{+1.37}$ & 2.95$_{-1.26}^{+1.24}$ & 4.87$_{-1.28}^{+1.38}$ & 4.09$_{-2.09}^{+2.35}$ & 2.15$_{-0.48}^{+1.21}$ & 5.41$_{-1.69}^{+1.52}$ & 3.56$_{-1.64}^{+1.28}$ & 2.72$_{-1.14}^{+1.63}$ & 4.06$_{-1.50}^{+0.79}$ & 4.18$_{-2.12}^{+1.03}$ & -- & -- \\ [1.5ex]
    
                \hline
    		\hline
    	\end{tabular}%
            }
            \tablefoot{
            Medians and 16-84 percentile intervals of the $n_{1.5\mu{\text{m}}}$ as a function of stellar mass in four redshift bins. Values are computed when at least 10 galaxies are available in the mass-redshift bin. The first column identifies the stellar mass bin investigated, then, grouped by three, are shown the values for the population as a whole (ALL), just the star-forming galaxies (SF), and just the quiescent (Q) corresponding to the redshift bin indicated above.
            }
        \end{table*}

        \begin{table*}
    	\centering
    	\caption{Same as Table \ref{tab:med_nir} but for $n_{0.5\mu{\text{m}}}$}
    	\label{tab:med_opt}
    	{\tiny
            \begin{tabular}{P{2cm}|P{0.75cm}P{0.75cm}P{1cm}|P{0.75cm}P{0.75cm}P{1cm}|P{0.75cm}P{0.75cm}P{1cm}|P{0.75cm}P{0.75cm}P{1cm}}
    		\hline
        	{} & \multicolumn{3}{c|}{$0.5\leq z<1$} & \multicolumn{3}{c|}{$1\leq z<1.5$} & \multicolumn{3}{c|}{$1.5\leq z<2$} & \multicolumn{3}{c}{$2\leq z<2.5$} \\
                \hline
                \rule{0pt}{0.3cm}
                Mass bin & ALL & SF & Q & ALL & SF & Q & ALL & SF & Q & ALL & SF & Q \\[0.4ex]
                \hline
                \rule{0pt}{0.4cm}
                $[10^{9.50}-10^{9.75}$) & 1.10$_{-0.54}^{+0.89}$ & 1.06$_{-0.52}^{+0.85}$ & 1.71$_{-0.86}^{+1.18}$ & 1.14$_{-0.52}^{+0.93}$ & 1.14$_{-0.52}^{+0.92}$ & 1.43$_{-0.46}^{+2.16}$ & 1.30$_{-0.60}^{+1.02}$ & 1.30$_{-0.61}^{+1.02}$ & -- & 1.32$_{-0.69}^{+1.39}$ & 1.31$_{-0.69}^{+1.39}$ & -- \\ [1.5ex]
                $[10^{9.75}-10^{10.00}$) & 1.01$_{-0.46}^{+0.93}$ & 0.95$_{-0.42}^{+0.80}$ & 1.93$_{-0.93}^{+1.29}$ & 1.03$_{-0.51}^{+0.94}$ & 1.01$_{-0.50}^{+0.93}$ & 1.55$_{-0.40}^{+1.69}$ & 1.24$_{-0.63}^{+1.22}$ & 1.23$_{-0.62}^{+1.20}$ & -- & 1.41$_{-0.81}^{+1.08}$ & 1.41$_{-0.81}^{+1.08}$ & -- \\ [1.5ex]
                $[10^{10.00}-10^{10.25}$) & 1.03$_{-0.45}^{+1.18}$ & 0.92$_{-0.37}^{+0.77}$ & 2.28$_{-0.96}^{+1.40}$ & 1.07$_{-0.52}^{+1.29}$ & 1.00$_{-0.48}^{+1.11}$ & 2.63$_{-1.26}^{+1.81}$ & 1.10$_{-0.56}^{+1.13}$ & 1.07$_{-0.54}^{+1.04}$ & 2.32$_{-0.71}^{+2.15}$ & 1.29$_{-0.66}^{+1.36}$ & 1.28$_{-0.65}^{+1.35}$ & -- \\ [1.5ex]
                $[10^{10.25}-10^{10.50}$) & 1.27$_{-0.66}^{+1.70}$ & 0.98$_{-0.44}^{+0.75}$ & 2.66$_{-1.08}^{+1.36}$ & 1.13$_{-0.65}^{+1.67}$ & 0.88$_{-0.44}^{+1.06}$ & 2.75$_{-1.28}^{+2.16}$ & 1.09$_{-0.54}^{+1.69}$ & 0.96$_{-0.44}^{+1.33}$ & 2.82$_{-1.23}^{+2.03}$ & 1.21$_{-0.67}^{+1.55}$ & 1.15$_{-0.62}^{+1.37}$ & 2.76$_{-1.36}^{+3.32}$ \\ [1.5ex]
                $[10^{10.50}-10^{10.75}$) & 1.85$_{-1.13}^{+1.84}$ & 1.12$_{-0.55}^{+1.52}$ & 3.08$_{-1.12}^{+1.25}$ & 1.34$_{-0.76}^{+1.78}$ & 1.01$_{-0.51}^{+1.14}$ & 2.58$_{-1.07}^{+1.58}$ & 1.70$_{-1.10}^{+1.99}$ & 0.98$_{-0.50}^{+1.53}$ & 2.92$_{-1.17}^{+1.45}$ & 1.60$_{-0.96}^{+2.46}$ & 1.10$_{-0.52}^{+1.62}$ & 3.52$_{-1.76}^{+1.88}$ \\ [1.5ex]
                $[10^{10.75}-10^{11.00}$) & 2.59$_{-1.28}^{+1.90}$ & 1.74$_{-0.84}^{+1.84}$ & 3.54$_{-1.24}^{+1.53}$ & 2.05$_{-1.38}^{+2.09}$ & 1.26$_{-0.69}^{+1.73}$ & 3.22$_{-1.14}^{+1.63}$ & 2.20$_{-1.40}^{+1.79}$ & 1.18$_{-0.62}^{+2.07}$ & 2.70$_{-0.93}^{+1.82}$ & 1.81$_{-1.15}^{+1.97}$ & 1.13$_{-0.67}^{+2.15}$ & 2.47$_{-1.12}^{+2.42}$ \\ [1.5ex]
                $[10^{11.00}-10^{11.25}$) & 3.45$_{-1.76}^{+1.72}$ & 2.76$_{-1.48}^{+1.72}$ & 4.33$_{-1.74}^{+1.10}$ & 2.46$_{-1.33}^{+2.08}$ & 1.76$_{-1.03}^{+1.46}$ & 3.29$_{-1.31}^{+1.64}$ & 2.51$_{-1.39}^{+1.85}$ & 1.64$_{-0.92}^{+1.83}$ & 3.23$_{-1.24}^{+1.40}$ & 2.40$_{-1.14}^{+2.43}$ & 2.19$_{-1.39}^{+1.73}$ & 2.97$_{-1.04}^{+2.27}$ \\ [1.5ex]
                $[10^{11.25}-10^{11.50}$) & 4.29$_{-1.90}^{+1.54}$ & 3.05$_{-1.37}^{+2.69}$ & 4.74$_{-1.60}^{+1.21}$ & 3.04$_{-1.73}^{+2.32}$ & 2.39$_{-1.51}^{+1.60}$ & 4.60$_{-1.68}^{+0.92}$ & 2.88$_{-1.55}^{+1.21}$ & 2.32$_{-1.09}^{+1.57}$ & 3.53$_{-1.25}^{+1.03}$ & 2.38$_{-1.24}^{+1.74}$ & 1.77$_{-0.91}^{+1.64}$ & 3.35$_{-1.14}^{+1.01}$ \\ [1.5ex]
                \hline
    		\hline
    	\end{tabular}%
            }
        \end{table*}

\end{appendix}

%
%

%
%
%
%
%
%
%
%
%

\end{document}